# COMMISSIONING AND FIRST RESULTS OF THE ELECTRON BEAM PROFILER IN THE MAIN INJECTOR AT FERMILAB*

R. Thurman-Keup†, M. Alvarez, J. Fitzgerald, C. Lundberg, P. Prieto, J. Zagel,
FNAL, Batavia, IL, 60510, USA
W. Blokland, ORNL, Oak Ridge, TN, 37831, USA

*Abstract*

The planned neutrino program at Fermilab requires large proton beam intensities in excess of 2 MW. Measuring the transverse profiles of these high intensity beams is challenging and often depends on non-invasive techniques. One such technique involves measuring the deflection of a probe beam of electrons with a trajectory perpendicular to the proton beam. A device such as this is already in use at the Spallation Neutron Source at ORNL and a similar device has been installed in the Main Injector at Fermilab. Commissioning of the device is in progress with the goal of having it operational by the end of the year. The status of the commissioning and initial results will be presented.

## INTRODUCTION

Traditional techniques for measuring the transverse profile of proton beams typically involve the insertion of an object into the path of the proton beam. Flying wires for instance in the case of circulating beams, or secondary emission devices for single pass beamlines. With increasing intensities, these techniques become riskier both for the device and the radioactivation budget of the accelerator. Various alternatives exist including ionization profile monitors, gas fluorescence monitors, and the subject of this report, electron beam profile monitors.

The concept of a probe beam of charged particles to determine a charge distribution has been around since at least the early 1970's [1-3]. A number of conceptual and experimental devices have been associated with accelerators around the world [4-8]. An operational device is presently in the accumulator ring at SNS [9].

An Electron Beam Profiler (EBP) has been constructed at Fermilab and installed in the Main Injector (MI) [10]. The MI is a proton synchrotron that can accelerate protons from 8 GeV to 120 GeV. The protons are bunched at 53 MHz with a typical rms bunch length of 1-2 ns. In this paper, we discuss the design and installation of the EBP and present some initial measurements.

## THEORY

The principle behind the EBP is electromagnetic deflection of the probe beam by the target beam under study (Fig. 1). If one assumes a target beam with $\gamma \gg 1$, no magnetic field, and $\rho \neq \rho(z)$, then the force on a probe particle is [11]

$$\vec{F}(\vec{r}) \propto \int d^2\vec{r}'\, \rho(\vec{r}')\, \frac{(\vec{r}-\vec{r}')}{|\vec{r}-\vec{r}'|^2}$$

and the change in momentum is

$$\Delta\vec{p} = \int_{-\infty}^{\infty} dt\, \vec{F}(\vec{r}(t))$$

For small deflections, $\vec{r} \approx \{b, vt\}$, and the change in momentum is

$$\Delta\vec{p} \propto \int_{-\infty}^{\infty} dx'\int_{-\infty}^{\infty} dy'\, \rho(x', y') \int_{-\infty}^{\infty} dt\, \frac{\{b-x', vt-y'\}}{(b-x')^2 + (vt-y')^2}$$

where {} indicates a vector. For small deflections, $\vec{p} \approx \{0, p\}$ and the deflection is $\theta \approx \frac{|\Delta\vec{p}|}{|\vec{p}|}$. The integral over time can be written as $\text{sgn}(b-x')$ leading to an equation for the deflection

$$\theta(b) \propto \int_{-\infty}^{\infty} dx' \int_{-\infty}^{\infty} dy'\, \rho(x', y')\, \text{sgn}(b-x')$$

where $\text{sgn}(x) = -1$ for $x < 0$ and $+1$ for $x \geq 0$.

If one takes the derivative of $\theta(b)$ with respect to $b$, the sgn function becomes $\delta(b-x')$ leading to

$$\frac{d\theta(b)}{db} \propto \int_{-\infty}^{\infty} dy'\, \rho(b, y')$$

which is the profile of the charge distribution of the beam. Thus, for a gaussian beam, this would be a gaussian distribution and the original deflection angle would be the error function, $\text{erf}(b)$.

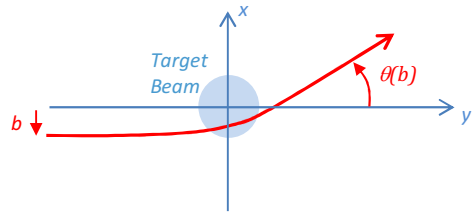

Figure 1: Probe beam deflection (red) for some impact parameter *b*.

## EXPERIMENTAL TECHNIQUE

To obtain $\theta(b)$, one needs to measure the deflection for a range of impact parameters. This can be accomplished in a single shot by sweeping the electron beam through the proton beam provided the sweep time is much smaller than the r.m.s. bunch length of the proton beam to avoid coupling the longitudinal and transverse distributions. In the main injector, this would be challenging considering

---

its bunch length is 1-2 ns. The electron beam can also be stepped through the proton beam while recording the deflection at each step as demonstrated in Fig. 2 [12].

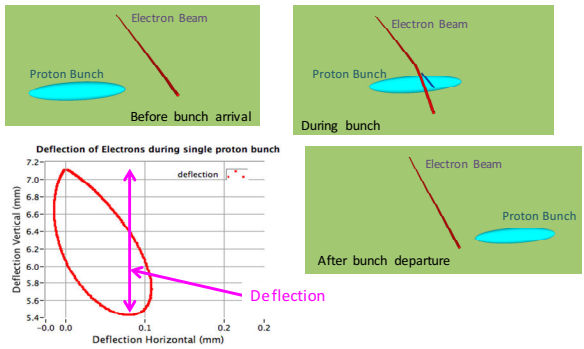

Figure 2: Trajectory followed by a stationary electron beam as the proton bunch passes by. There is some deflection along the proton beam direction due to the magnetic field of the proton beam, but it is much smaller than the deflection transverse to the proton beam.

The method chosen for the MI implementation is a variation of the slow stepping. Instead of a stationary electron beam at each step, the beam is swept along the direction of the proton beam producing an approximate longitudinal distribution (Fig. 3). This technique has the potential to allow longitudinal slicing of the transverse profile assuming the longitudinal distribution either remains constant over the series of impact parameter measurements or can be corrected for synchrotron motion.

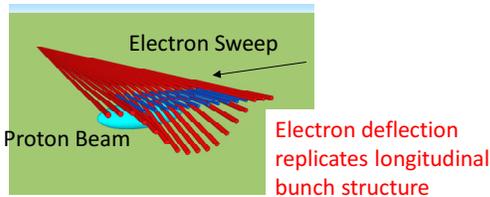

Figure 3: The electron beam is swept along the direction of the proton beam with a sweep time comparable to the proton bunch length. This records the deflection as a function of longitudinal position. A series of these sweeps is collected at different impact parameters to obtain $\theta(b)$.

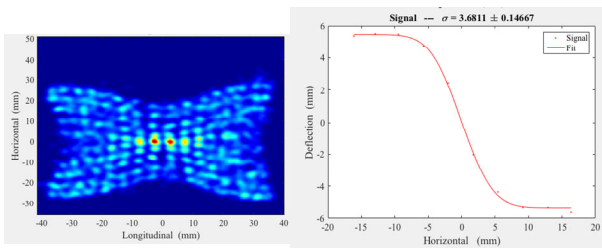

Figure 4: Simulated image of successive electron beam sweeps along the proton beam direction. The sweeps near the center are difficult to separate and may need to be split across multiple camera frames.

Simulations of the deflection are shown in Fig. 4. Here successive sweeps along the proton direction at different impact parameters are displayed in the same image. Each simulated electron produces a Gaussian spot with an rms of 1 mm. The simulation was done for injection parameters of 3 mm horizontal proton beam size, and 2 ns bunch length. One can see that the central deflections may overlap each other. Problems such as these must be overcome through, for example, timing shifts or interleaving across multiple camera frames.

## APPARATUS

The device that was constructed for the MI consists of an EGH-6210 electron gun from Kimball Physics, followed by a cylindrical, parallel-plate electrostatic deflector, and finally a phosphor screen acting as the beam dump (Fig. 5).

The gun (Fig. 6) is a 60 keV, 6 mA, thermionic gun with a LaB$_6$ cathode, that can be gated from 2 μs to DC at a 1 kHz rate. The gun contains a focusing solenoid and four independent magnet poles for steering/focusing. The minimum working spot size is <100 μm. The electrostatic deflector (Fig. 6) contains 4 cylindrical plates that are 15 cm long and separated by ~2.5 cm. Following the electrostatic deflector is the intersection with the proton beamline. There is a pneumatic actuator at this point with a stainless-steel mirror for generating optical transition radiation (OTR) to be used in calibrating the electron beam.

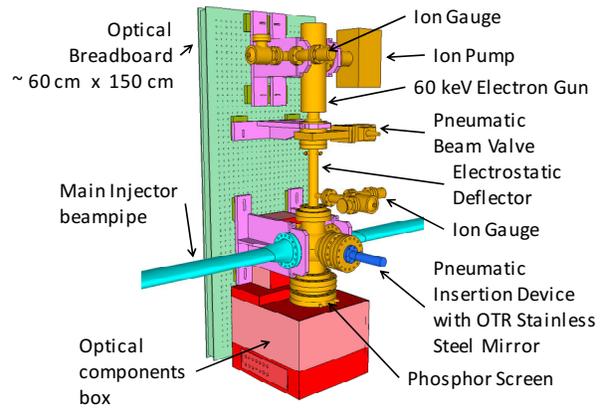

Figure 5: Model of the EBP showing the main components.

After the proton beam intersection, there is a phosphor screen from Beam Imaging Systems (Fig. 6). It is composed of P47 (Y$_2$SiO$_5$:Ce3+) with an emission wavelength of 400 nm, a decay time of ~60 ns and a quantum yield of 0.055 photons/eV/electron. The phosphor screen has a thin conductive coating with a drain wire attached. Both the OTR and the phosphor screen are imaged by a single intensified camera system (Fig. 7). The source is chosen by a mirror on a moving stage. Each source traverses a two-lens system plus optional neutral density filters or polarizers before entering the image intensifier (Hamamatsu V6887U-02). The output of the intensifier is imaged by a COHU CCD camera with C-mount lens. This setup will likely change in favour of a CID camera from Thermo-electron (now Thermo Scientific) fiber-optically coupled to the intensifier through a fiber-optic taper to improve light collection.

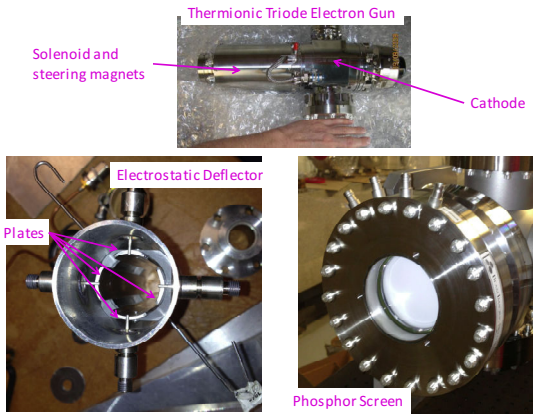

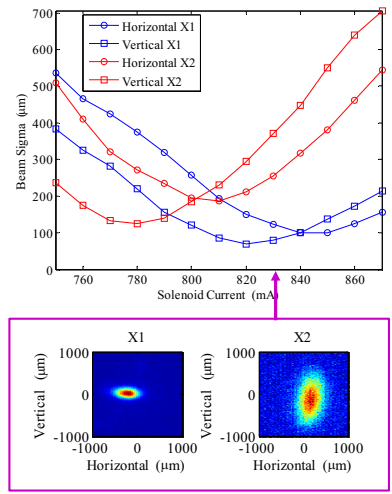

Figure 6: Top) Commercial electron gun. Left) Inside view of the electrostatic deflector showing the cylindrical parallel plates. Right) Phosphor screen mounted to an 8 in conflat flange with viewport. A drain wire is attached between the screen and one of the SHV connectors.

Figure 8: Horizontal and vertical rms beam sizes at the first (blue) and second (red) crosses in the test stand. The measurements are from OTR taken at ~50 keV and 1 mA beam current onto the stainless-steel mirrors.

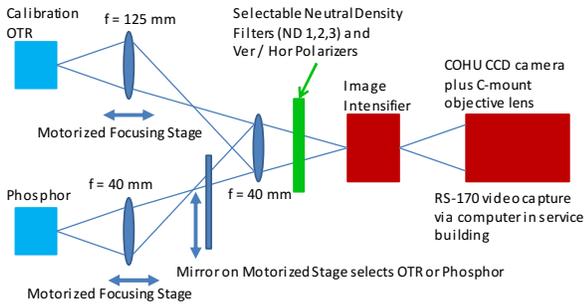

Figure 7: Conceptual layout of the optical paths followed by the OTR light and the phosphor screen light. Of the two lenses in each path, the second one is shared.

## TEST STAND RESULTS

A test stand was setup to measure beam characteristics of the electron gun. It consisted of a pair of OTR screens used to measure the spot size and divergence to verify the manufacturer's specifications and for use in the simulation. The beam measurements were carried out using the solenoidal magnet in the gun to focus the beam at the first screen, allowing a measurement of the emittance of the electron beam (Fig. 8). Although these measurements were done at 50 keV, the intensity of the MI beam requires an electron energy of only 10-15 keV.

## INSTALLATION

The EBP was installed in the MI during the 2014-2016 maintenance periods (Fig. 9). The location is near the end of a straight section just upstream of a horizontal focussing quadrupole. The expected horizontal rms beam size at this location is expected to be 1-3 mm. Because of the proximity to the MI magnet busses, the entire EBP beamline was wrapped in three layers of mumetal, mostly eliminating electron beam movement due to bus currents.

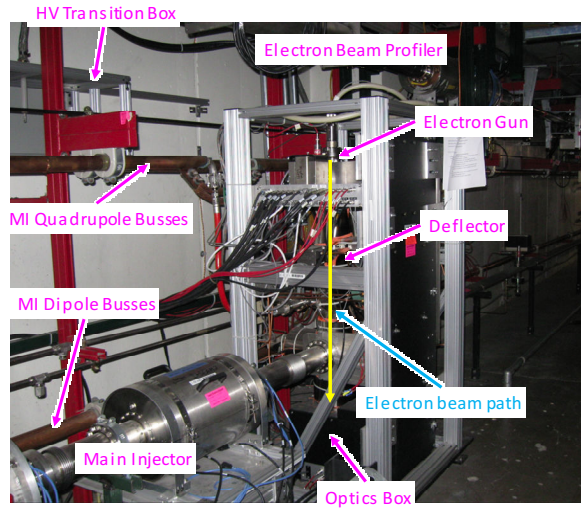

Figure 9: EBP installed in the end of a straight section in the MI. One can see the close proximity to the magnet busses.

## RESULTS

Initial measurements of transverse profiles of the MI beam have been made with stationary electron beams. Figure 10 shows the deflection of the stationary electron beam as the proton bunches pass by the electron beam. The bright spot is the undeflected electron beam as seen in the bottom pictures where there is no proton beam. There is still a bright spot in the upper pictures due to the gaps between the proton bunches when the electron beam is not deflected. These images are taken just after injection into the MI at 8 GeV. The expected horizontal beam size is about 3 mm at this location at injection.

Using the amount of deflection in the images, a plot of deflection vs. impact parameter can be formed from which to extract the proton beam size (Fig. 11). The measured rms horizontal beam size at injection is about 3.7 mm.

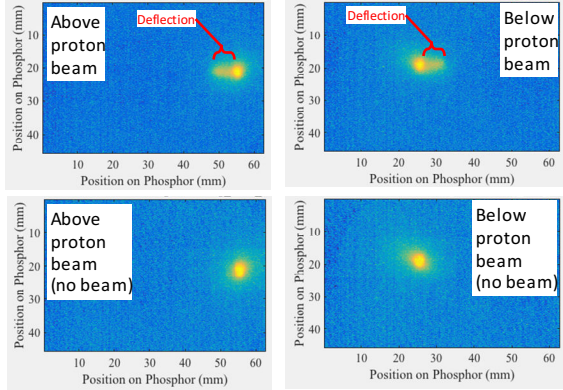

Figure 10: Deflection of the electron beam (top) for cases of the electron beam above and below the proton beam. The bottom images are without the proton beam.

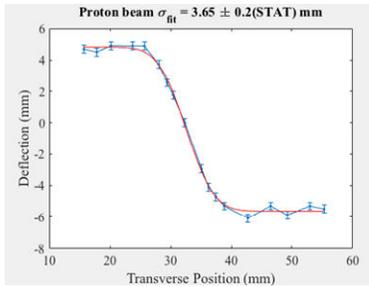

Figure 11: Electron beam deflection as a function of impact parameter with the proton beam. The uncertainties are just an estimate of how well one can determine the peak deflection from the images in Fig. 10.

Measurements were also taken at extraction from MI at 120 GeV and near transition crossing at about 19 GeV (Fig. 12). The expected rms beam size at extraction is about 1 mm. The other features to note in these images are the amount of deflection and the intensity of the deflected part of the electron beam. The bunch length of the protons is shorter at extraction than at injection and is particularly short near transition. There are two consequences of these facts: the charge density is higher at extraction and transition which produces larger deflections, and the shorter bunches result in a smaller proportion of deflected beam. This is consistent with what is seen in the images in Fig. 12.

To study the intended fast deflection along the proton beam line (Fig. 3), the deflector was tested using a FET-based HV pulser, without the proton beam present. An electron beam streak is shown in Fig. 13. This image contains both the primary sweep (~20 ns), and the return sweep which is about 5 times slower. Thus, the intensity of just the primary would be significantly less.

## SUMMARY

An electron beam profiler has been built and installed in the MI at Fermilab and has been used to make some measurements of the horizontal proton beam size. The measurements are in fairly good agreement with the expected values. There are several repairs and improvements that need to be made. Some of these are in progress during the current maintenance period. The phosphor screen will be replaced and the camera system will be modified to be more radiation and noise tolerant.

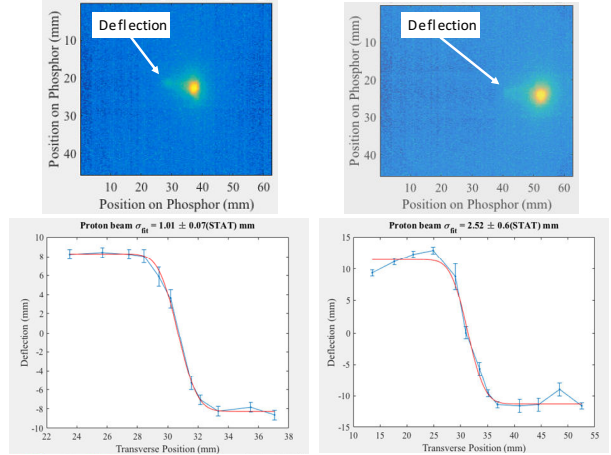

Figure 12: Deflection vs. impact parameter for extraction (left) and transition crossing (right). The bunch length at transition is shorter than extraction, so the deflection is larger due to increased charge density. The intensity is also slightly less, since the electron beam spends less time being deflected.

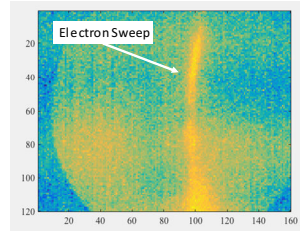

Figure 13: Electron beam streak using the deflecting plates. The image contains both the primary and return streaks. The curvature is due to the curved deflecting plates.

## ACKNOWLEDGEMENTS

The authors would like to acknowledge the help of the MI, Mechanical, Electrical, and Instrumentation departments for all their assistance in the construction and installation of this device. This manuscript has been authored by Fermi Research Alliance, LLC under Contract No. DE-AC02-07CH11359 with the U.S. Department of Energy, Office of Science, Office of High Energy Physics. The United States Government retains and the publisher, by accepting the article for publication, acknowledges that the United States Government retains a non-exclusive, paid-up, irrevocable, world-wide license to publish or reproduce the published form of this manuscript, or allow others to do so, for United States Government purposes.

## REFERENCES

[1] Paul D. Goldan, "Collisionless sheath – An experimental investigation", *Phys. Fluids,* vol. 13, p. 1055, 1970.

[2] C. H. Stallings, "Electron beam as a method of finding the potential distribution in a cylindrically symmetric plasma", *J. Appl. Phys.*, vol. 42, p. 2831, 1971.

[3] C. W. Mendel Jr., "Apparatus for measuring rapidly varying electric fields in plasmas", *Rev. Sci. Instrum.*, vol. 46, p. 847, 1975.

[4] J. Shiloh et al., "Electron beam probe for charge neutralization studies of heavy ion beams", *Rev. Sci. Instrum.*, vol. 54, p. 46, 1983.

[5] V. Shestak et al., "Electron beam probe for ion beam diagnostics", TRIUMF, Rep. TRI-DN-87-36, 1987.

[6] P. Gross et al., "An electron beam probe for ion beam diagnosis", in *Proc. European Particle Accelerator Conference 1990*, Nice, France, June 1990, p. 806.

[7] J. Bosser et al., "Transverse profile monitor using ion probe beams", *Nucl. Instrum. Methods Phys. Res. A*, vol. 484, p. 1, 2002.

[8] P. V. Logatchov et al., "Non-destructive singlepass monitor of longitudinal charge distribution in an ultrarelativistic electron bunch", in *Proc. Particle Accelerator Conference 1999*, New York, USA, March 1999.

[9] W. Blokland and S. Cousineau, "A non-destructive profile monitor for high intensity beams", in *Proc. Particle Accelerator Conference 2011*, New York, USA, March 2011.

[10] R. Thurman-Keup et al., "Installation status of the electron beam profiler for the Fermilab Main Injector", in *Proc. International Beam Instrumentation Conference 2015*, Melbourne, Australia, September 2015.

[11] A. Aleksandrov et al., "Feasibility study of using an electron beam for profile measurements in the SNS accumulator ring", in *Proc. Particle Accelerator Conference 2005*, Knoxville, USA, May 2005, pp. 2586 – 2588.

[12] W. Blokland, "Collaboration report on e-beam scanner for Project X beam instrumentation", unpublished, October 2011.